\documentclass[fleqn,10pt]{wlscirep}
\usepackage[utf8]{inputenc}
\usepackage[T1]{fontenc}
\usepackage{float}
\title{Analyzing the progress of Indian states chasing sustainable development goals using complex network framework}

\author[1]{Hrishidev Unni}
\author[2,3]{Rubal Rathi}
\author[3]{Sangita Dutta Gupta}
\author[1,*]{Anirban Chakraborti}

\affil[1]{School of Computational \& Integrative Sciences, Jawaharlal Nehru University, New Delhi-110067, India}
\affil[2]{Indian School of Hospitality, Sector-83, Gurugram-122004, Haryana, India}
\affil[3]{BML Munjal University, Sidhrawali, Gurugram-122413, Haryana, India}

\affil[*]{anirban@jnu.ac.in}

\begin{abstract}
The Sustainable Development Goals (SDGs) offer a critical global framework for addressing challenges like poverty, inequality, climate change, etc. They encourage a holistic approach integrating economic growth, social inclusion, and environmental sustainability to create a better future. We aim to examine India’s responsibility in achieving the SDGs by recognizing the contributions of its diverse states in the federal structure of governance. As the nodal agency in India, the NITI Aayog’s existing SDG index, using various socioeconomic indicators to determine the performance across different goals, serves as a foundation for assessing each state’s progress. Building on the seminal works of Hidalgo and Hausmann (2009) and Tachhella et al. (2012), which introduced the economic complexity/fitness index, Sciarra et al. (2020) proposed the SDGs-Generalized Economic Complexity (GENEPY) framework to quantify “complexity” by computing “ranks for states” and “scores for goals”, treating them as part of a complex bipartite network. In this paper, we apply the SDGs-GENEPY, to evaluate the progress and evolution of Indian states and union territories over several years. This enables us to identify each state’s capacity (and rank) in achieving the SDGs. We can interpret these complexity scores as “centrality measures” of a complex bipartite network of the states and the goals. This enhances our understanding of the complex relationship between state capabilities and the achievability of SDGs within the Indian context and enables data-driven policy-making.
\end{abstract}
\begin{document}

\flushbottom
\maketitle
\thispagestyle{empty}

\section*{Introduction}

For a long time, economic discussions have centered around growth as a key development measure. However, relying solely on growth as an indicator has limitations and certainly does not capture the full picture of development. Many studies have emphasized the importance of a more holistic approach to measuring development rather than just Gross Domestic Product \cite{lewis1965review,schultz1963economic,schultz1971investment,schumacher1973small,stiglitz1998towards,stiglitz2018beyond}. However, there are always policy challenges in reconciling the holy trinity of equity, efficiency, and sustainability \cite{munasinghe2005role,fleurbaey2013beyond}. The concept of sustainable development offers a more comprehensive and balanced understanding of development compared to traditional models that focus mainly on economic growth. Unlike the traditional growth-centric approaches, sustainable development integrates social, environmental, and economic dimensions, recognizing that long-term prosperity requires more than just increased output \cite{un2023sdgreport}. Economic growth alone would certainly lead to the depletion of natural resources, environmental degradation, and social inequalities, which may undermine the ability of future generations to thrive. Sustainable development addresses these issues by promoting policies and practices that balance economic progress with environmental preservation and social well-being. It emphasizes responsible resource use, renewable energy, and reducing carbon footprints, ensuring that development today does not compromise the ability of future generations to meet their own needs. Furthermore, sustainable development prioritizes equity, aiming to improve the quality of life for all, especially marginalized groups, through access to education, healthcare, and equal opportunities. By integrating these diverse aspects, sustainable development provides a more inclusive and forward-looking framework that acknowledges the interconnectedness of economic, social, and environmental factors, making it a far better approach to measuring and achieving true development \cite{wb2023uncertainty}.

With the above context of sustainable development, the Millennium Development Goals (MDGs) were framed in September 2000 during the United Nations Millennium Summit. At this summit, world leaders from 189 countries adopted the United Nations Millennium Declaration. The MDGs consisted of eight global development goals aimed at addressing issues such as poverty, hunger, education, gender equality, child mortality, maternal health, combating diseases, environmental sustainability, and global partnership, with the initial target for achieving these goals set as 2015. The idea of the Sustainable Development Goals (SDGs) emerged from the need to build on the progress and address the limitations of the Millennium Development Goals (MDGs), which were set to expire in 2015. The concept of SDGs was first proposed at the United Nations Conference on Sustainable Development, known as the Rio+20 Summit, held in Rio de Janeiro, Brazil, in June 2012. At the Rio+20 Summit, 193 member states recognized that development must be sustainable, balancing social, economic, and environmental aspects. They produced the outcome document, titled “The Future We Want”, which called for the establishment of a set of universal goals that would address global challenges more holistically. These new goals were intended to continue the work of the MDGs but expand their scope to include environmental sustainability, economic inequality, and social inclusivity, ensuring that no one is left behind \cite{sachs2022crisis}. The United Nations formally adopted the Sustainable Development Goals (SDGs) in September 2015 during the UN Sustainable Development Summit in New York, with a global commitment to achieve these 17 goals by 2030. The concept of the SDGs emerged after several years of negotiations involving policymakers, NGOs, and experts. The primary goal was to establish a set of targets for member nations to address economic, political, and environmental challenges \cite{Sachs_Kroll_Lafortune_Fuller_Woelm_2021}. Each of the 17 SDGs is accompanied by multiple subgoals, all designed to benefit both people and the planet. The SDGs aim to eliminate poverty, promote a better future for all, protect the environment, and ensure a dignified life for individuals in the member countries \cite{xu2022assessing,gao2017finding}.

Several recent crises have significantly impaired the progress toward achieving the Sustainable Development Goals (SDGs):
COVID-19 Pandemic: The global pandemic, which began in early 2020, had far-reaching consequences on health, economies, education, and social systems. It led to widespread job losses, disruptions in healthcare services, school closures, and increased poverty. The economic setbacks caused by COVID-19 reversed years of progress, particularly on goals related to health (SDG 3), poverty (SDG 1), and education (SDG 4).
War in Ukraine: The ongoing conflict between Russia and Ukraine, which began in early 2022, has created major disruptions in global food and energy supply chains. Ukraine and Russia are significant producers of grains, fertilizers, and energy. The war has caused food insecurity, higher energy prices, and economic instability worldwide, negatively affecting progress on SDGs related to food security (SDG 2), energy (SDG 7), and peace (SDG 16).
Climate Change: The accelerating impacts of climate change, including extreme weather events such as floods, droughts, and wildfires, have worsened environmental degradation and increased vulnerability for many communities. Climate change has impacted efforts to achieve goals related to clean water (SDG 6), climate action (SDG 13), and life on land and below water (SDGs 14 and 15).
Global Economic Slowdowns: Various economic crises, driven by the pandemic, inflation, energy price shocks, and geopolitical tensions, have led to slower global growth and deepened inequalities. This has hindered progress in reducing inequality (SDG 10), eradicating poverty (SDG 1), and ensuring decent work and economic growth (SDG 8) \cite{imf2022cost}. These crises have significantly affected the advancement of member countries \cite{sdgreport2022}. As a result, policymakers have shifted their focus from the long- and medium-term goals of the SDGs to addressing immediate, pressing challenges. The need for fundamental SDG cooperation and clean energy has become more critical than ever in the current global context \cite{wang2022dynamic,biermann2022scientific}.

The Sustainable Development Goals (SDGs) are inherently complex and multi-faceted, which has led to the development of composite indices that offer more meaningful ways to assess their achievability. At the global level, SDG indices cover all 193 UN member states, with the Sustainable Development Solutions Network (SDSN) providing such indices since 2015. The SDSN’s latest 2022 report continues this initiative, offering a broad view of SDG progress across countries \cite{sachs2022crisis}. However, creating SDG indices at the subnational level remains a work in progress, especially in large, diverse nations like India \cite{cao2023spatio, wang2022study}. Given India’s federal structure, it is critical for individual states to actively participate in SDG advancement, necessitating a state-wise SDG index to help identify specific progress and areas needing focus.
In response to this need, NITI Aayog created the SDG India Index, marking the first government-led effort to track SDG progress at the subnational level. Designed to map the progress of each state and union territory, this index not only measures advancement but also fosters cooperative and competitive federalism to encourage further action on the SDGs (Economic Survey, 2021-22). NITI Aayog released the SDG Index annually from 2018 to 2021-22 and most recently in 2023-24, with the latest edition assessing performance across 16 SDGs, ranking states and monitoring their progress over time. SDG 17, focusing on global partnerships, was evaluated qualitatively. Notably, the first edition in 2018 excluded SDGs 12, 13, 14, and 17, leading the Observer Research Foundation (ORF) to create its own SDG Index in 2019, arguing that the omission of SDG 13 (Climate Action) left the original index incomplete. ORF’s index incorporated 14 of the 17 SDGs, providing a more comprehensive view of state progress in climate action and related goals \cite{ghosh2019sdg, sethi2022sdgs}.
While some studies have explored SDG interconnections in India, there remains a need for detailed analyses of state-level contributions toward SDG achievements. As each Indian state contributes uniquely to SDG progress, establishing a bipartite network that links states to their SDG achievements could facilitate systematic rankings of these contributions, enabling more granular assessments. This study seeks to address that gap by applying a complexity measure to calculate the SDG index, treating states and SDGs as nodes in a bipartite network to better understand each state’s capacity to achieve the SDGs.

Our approach draws on the pioneering work of Hidalgo and Hausmann (2009), who developed the economic complexity index to assess a country’s knowledge base based on its export data. Their method, called the “method of reflection,” uses an iterative approach to link product ubiquity with country diversification, resulting in scores that capture a nation’s growth potential through economic complexity \cite{hidalgo2009building}. In a later study, Tacchella et al. (2012) refined this method, suggesting that linear models could not fully capture complexity and advocating for a non-linear approach instead \cite{tacchella2012new}. Sciarra et al. (2020) built upon these insights, placing both methods within a network theory framework and redefining complexity as centrality scores within similarity matrices, which they found useful for exploring relationships in SDG achievement as well \cite{sciarra2020reconciling}. This study applies the SDGs-Generalized Economic Complexity (GENEPY) framework proposed by Sciarra et al.\cite{sciarra2021network} to analyze India’s SDG progress, facilitating a better understanding of state-level contributions and complexities within a broader, networked framework. By identifying these connections, we aim to enhance understanding of state capabilities and support more informed, data-driven policy-making for sustainable development in India.

\section*{Datasets}
The data used for this study was obtained from the SDG Dashboard provided by NITI Aayog. NITI The Aayog, as the nodal agency for SDGs in India, has released SDG reports for all Indian states and Union Territories for the years 2018, 2019, 2020-21, and 2023-24. Each of the reports comes with a data frame that consists of available scores ranging from 0-100 of 28 states and 8 union territories in 16 sustainable development goals. These scores are constructed by aggregating data from several indicators (62 indicators in 2018, 100 in 2019, 115 in 2020-21, and 113 in 2023-24). For comparability across various states and goals, these indicators were normalized by re-scaling them to a range between 0 and 100. Scores approaching 100 indicate that the state or Union Territory is nearing the target value set for an indicator, meaning they are closer to achieving the corresponding SDG. These normalized scores are then averaged to arrive at the score for a state in a particular goal. The 2018 dataset excludes goals 11,12 and 13 due to the unavailability of comparable data sources. In 2018 as well as the other years, the 14th goal “Life Below Water” is not considered for the analysis since it does not apply to states without shorelines which are in the majority. The 17th goal was also not included in the analysis since the NIF has not identified the relevant indicators for this goal.

\section*{Methodology}
To explore the relationships between SDGs and their status across Indian states, the dataset is extracted from the NITI Ayog SDG dashboard for the 4 years (iterations/versions) of SDG reports released. NITI Aayog measures the progress of a state $s$ in a particular SDG goal $g$ for a year $\tau$ through a set of indicators. For an indicator $k$ corresponding to a goal $g$, let $I_{sgk}(\tau)$ be the score of a state/UT $s$ in that indicator for a year $\tau$. These scores are calculated after normalizing and re-scaling the values of that indicator such that the states that reached a chosen target will be given 100 whereas the worst performing state receives a 0. By taking the average scores of all the indicators $k$ belonging to a goal $g$, we can calculate the scores of the states $s$ in different goals. $I_{sg}(\tau) = \sum_k I_{s g k}(\tau)/\nu^{k}_g(\tau)$ where $\nu^{k}_g(\tau)$ is the number of indicators present in goal $g$ for that year. NITI Aayog calculates the final composite index score of each state by taking the average of all the goal-specific performances applicable for the state (eg. goal 14 (which is about life below water) is not applicable in states/UTs without shoreline). 

This study considers SDGs and Indian states as a bipartite network with weighted links, where the weights correspond to the scores obtained. It further employs the framework by Sciarra et al.\cite{sciarra2020reconciling} called Generalized Economic Complexity (SDGs-GENEPY). The goal of this framework (as well as its predecessors) is to obtain two sets of related scores for both the states ($D_s(\tau)$) and the different SDGs ($C_g(\tau)$), where these scores are related to each other by an iterative relationship. The difference between the previously existing approaches was in articulating this relationship between the two scores. For this study, we adopt the following interpretation: "More complex states are capable of achieving more complex goals, and more complex goals are those that can only be achieved by the more complex states." This formalization, as outlined by Tachella et al., can be expressed as a non-linear iterative algorithm. 
$$
\left\{\begin{array}{ll}\widetilde{D_{s}}^{(n+1)}(\tau)={\sum }_{g}{I}_{sg}(\tau){C}_{g}^{(n)}(\tau),&{D}_{s}^{(n+1)}(\tau)=\frac{{\widetilde{{D}_{s}}}^{(n+1)}(\tau)}{\left({\sum }_{s}{\widetilde{{D}_{s}}}^{(n+1)}(\tau)\right)/\nu^s(\tau)}; \\ \widetilde{C_{g}}^{(n+1)}(\tau)=\frac{1}{{\sum }_{s}{I}_{sg}(\tau)\frac{1}{{D}_{s}^{(n)}(\tau)}},&{C}_{g}^{(n+1)}(\tau)=\frac{{\widetilde{{C}_{g}}}^{(n+1)}(\tau)}{\left({\sum }_{g}{\widetilde{{C}_{g}}}^{(n+1)}(\tau)\right)/\nu^g(\tau)};\end{array}\right.
$$
Sciarra et. al showed that one could approximate this to an eigen-problem by linearizing the system of equations, enabling us to interpret the two scores as eigenvectors of a similarity matrix among the states as well as among the goals. 
We can interpret the original data as a weighted bipartite network where the SDGs and states constitute the two sets of nodes. This lets us understand the scores as centrality metrics for ranking the importance of nodes within the network.

\subsection*{The SDGs-GENEPY framework}
\begin{figure}[h]
\centering
\includegraphics[width=.55\linewidth]{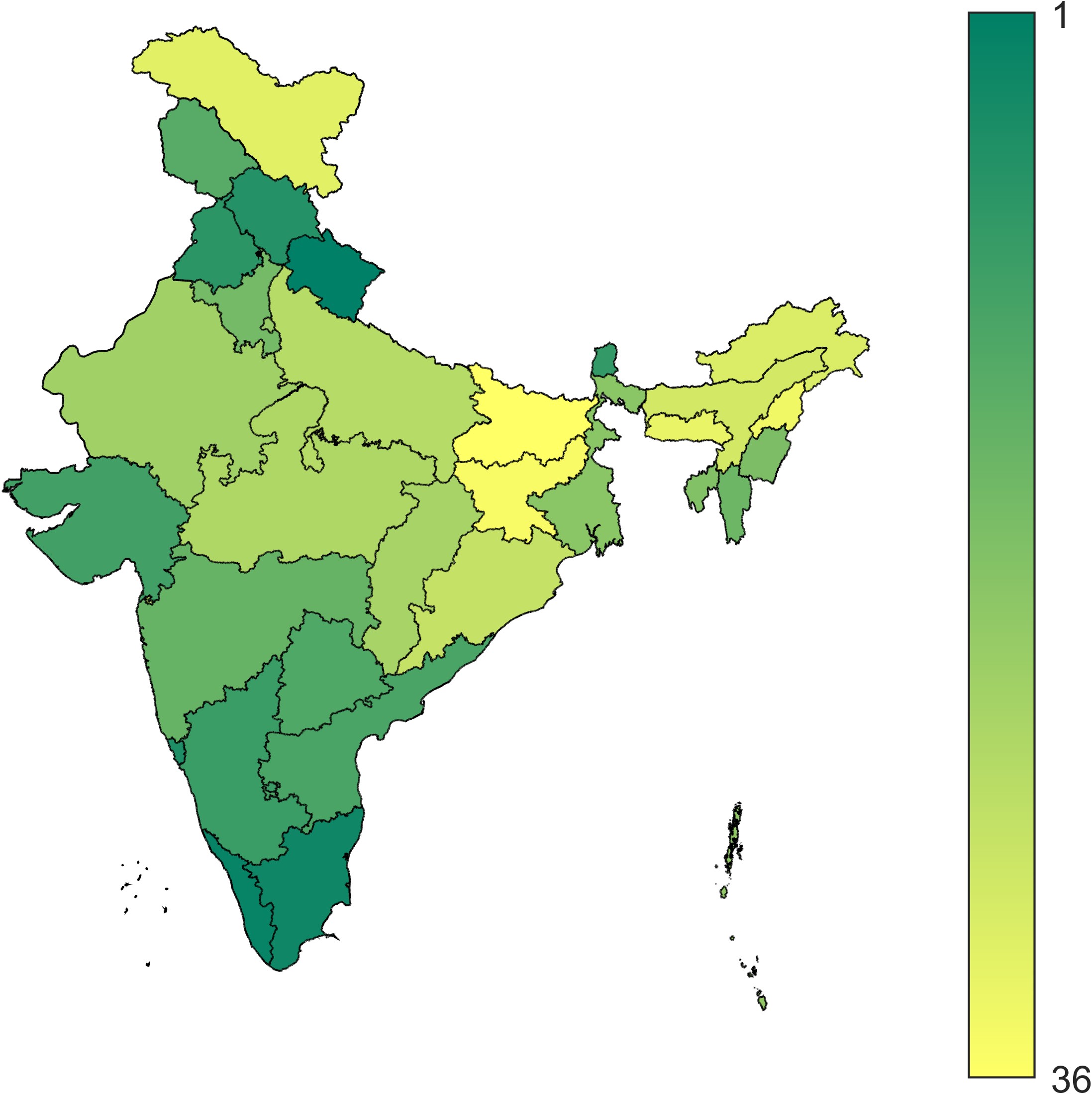}
\includegraphics[width=.43\linewidth]{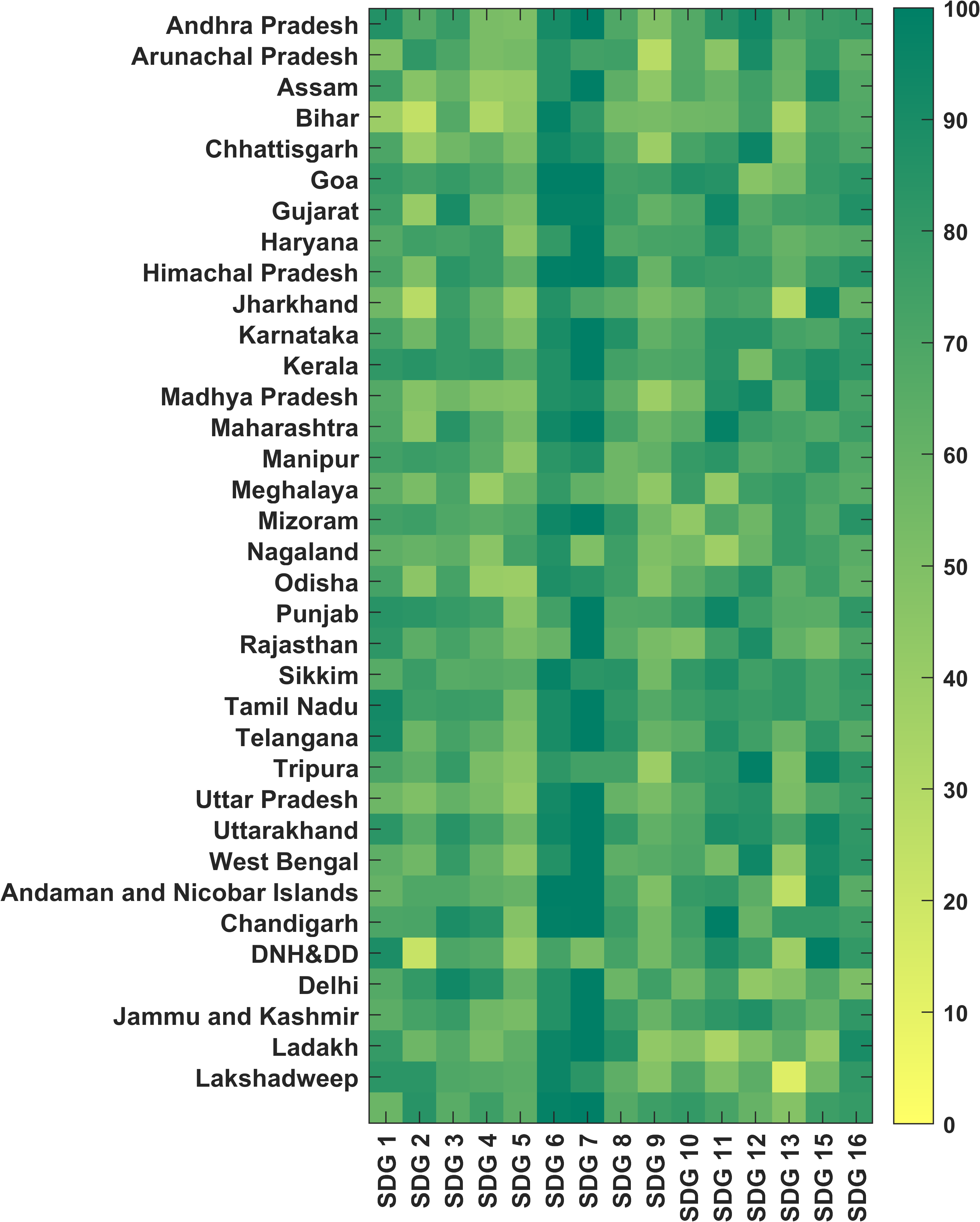}
\caption{\textbf{NITI Aayog dataset.} In sub-figure A, The data matrix ($I_{sg}(2024)$) consisting of 36 states + union territories and their respective scores in each of the sustainable development goals is visualised by mapping the values to a colourmap. The higher values are towards green and the lower values are towards yellow. In sub-figure B, Indian states and union territories are coloured according to their rankings based on the sums of individual SDG scores which correspond to both $k_s(2024)$ and the NITI Aayog's average scores. 
}
\label{fig:Fig1}
\end{figure}

We start with the matrix $\textbf{I} (\tau)$ for a year $\tau$ containing $\nu^s(\tau)$ rows corresponding to the states and union territories, and $\nu^g(\tau)$ columns corresponding to the SDGs that are present in that year. Each element $I_{sg}(\tau)$ corresponds to the score achieved by state $s$ on goal $g$. This matrix is visualized by mapping the values to a colourmap for Figure \ref{fig:Fig1}A.
The states + union territories and their respective scores in each sustainable development goal can be considered a bipartite network where the edge weights are given by the scores $I_{sg}(\tau)$. It is immediately clear that the NITI Aayog's composite scores for a state $s$ is given by the weighted degree centrality of the nodes $\sum_g{I_{sg}(\tau)}$. A subset of the network with only 6 states sampled across the ranking ladder is shown in Figure \ref{fig:Fig1}B. The chosen states are - Kerala(KL), Chandigarh(CH), Haryana(HR), Mizoram(MZ), Jharkhand (JH) and Bihar(BR).

The goal of this framework is to obtain two sets of related scores for both the states ($D_s(\tau)$) and goals ($C_g(\tau)$) corresponding to the centralities of the nodes in the bipartite network given by $I_{sg}(\tau)$ we defined above. These scores are motivated by the iterative relationship that the more complex states are the ones achieving more complex goals, and the more complex goals are the ones that are achieved by only the more complex states. 
To start we can define two vectors $k_s(\tau)$ and $k_g^{\prime}(\tau)$ as the column-wise and an adjusted row-wise sum of $I_{sg}(\tau)$ respectively.
$$
k_g^{\prime}(\tau)=\sum_s I_{s g}(\tau)/ k_s(\tau), k_s(\tau)=\sum_s I_{s g}(\tau)
$$ %Put them in 2 lines instead of 1. 

$k_s(\tau)$ is the total score of the state $s$ and $k_g^{\prime}(\tau)$ is the $g$ goal’s score adjusted for the relative performances of states achieving them. $k_s(\tau)$ directly corresponds to the NITI Ayog's SDG index since it is the average of the scores, and $k_s(\tau)$ is the sum. For all purposes, we will use $k_s$ to represent the SDG index scores given by NITI Ayog. Using these two vectors along with $\textbf{I}(\tau)$ we can compute a matrix $N(\tau)$ whose elements are given by
$$
N_{s g}(\tau)=\frac{I_{s g}(\tau)}{k_s(\tau) k_g^{\prime}(\tau)} .
$$

\begin{figure}[h]
\centering
\includegraphics[width=.9\linewidth]{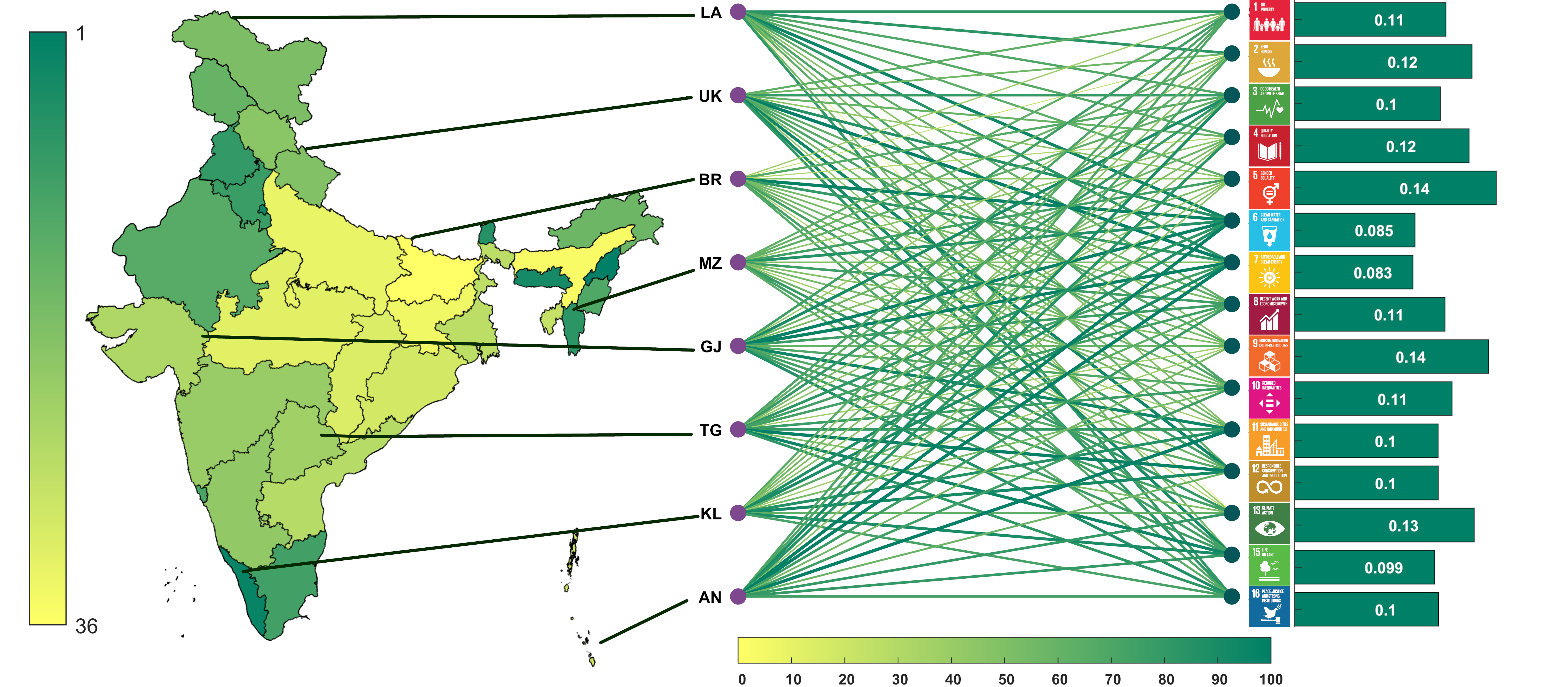}
\caption{\textbf{SDGs-GENEPY framework.} A subset of the underlying weighted bipartite network with 8 sample states (Andaman and Nicobar(AN), Kerala(KL), Telangana (TG), Gujarat(GJ), Mizoram(MZ), Jharkhand(JH), Bihar(BR) and Ladak(LA)) is visualized in the sub-figure B. The edges are coloured and drawn according to the weight of the links ($I_{s g}(2024)$) where the thicker and greener edges represent a stronger performance of a state in that particular goal. In sub-figure A, Indian states and union territories are coloured according to their rankings based on complexity scores ($D_s(2024)$). In sub-figure C the weights of the goals ($W_g(2024)$) obtained are shown as a horizontal bar chart.}
\label{fig:Fig2}
\end{figure}

This matrix formalizes the rationale given in the iterative framework explained before and provides a symmetric representation of the bipartite system. From here we can calculate the vector $D_s(\tau)$ as the principal eigenvector of the symmetric matrix $U(\tau)$ given by. 

$$
U_{s s^*}(\tau)=\mathbf{N(\tau) N}^{\prime}(\tau)=\sum_g \frac{I_{s g}(\tau) I_{s^* g}(\tau)}{k_s(\tau) k_{s^*}(\tau)\left(k_g^{\prime}(\tau)\right)^2}
$$

And our score vector $C_g(\tau)$ can be computed as the principal eigenvector of the conjugate symmetric matrix $V$ given by 

$$
V_{g g^*}(\tau)=\mathbf{N}^{\prime}(\tau) \mathbf{N}(\tau)=\sum_s \frac{I_{s g}(\tau) I_{s g^*}(\tau)}{k_s(\tau)^2 k_g(\tau)^{\prime} k_{g^*}(\tau)^{\prime}}
$$

\begin{figure}[H]
\centering
\includegraphics[width=.9\linewidth]{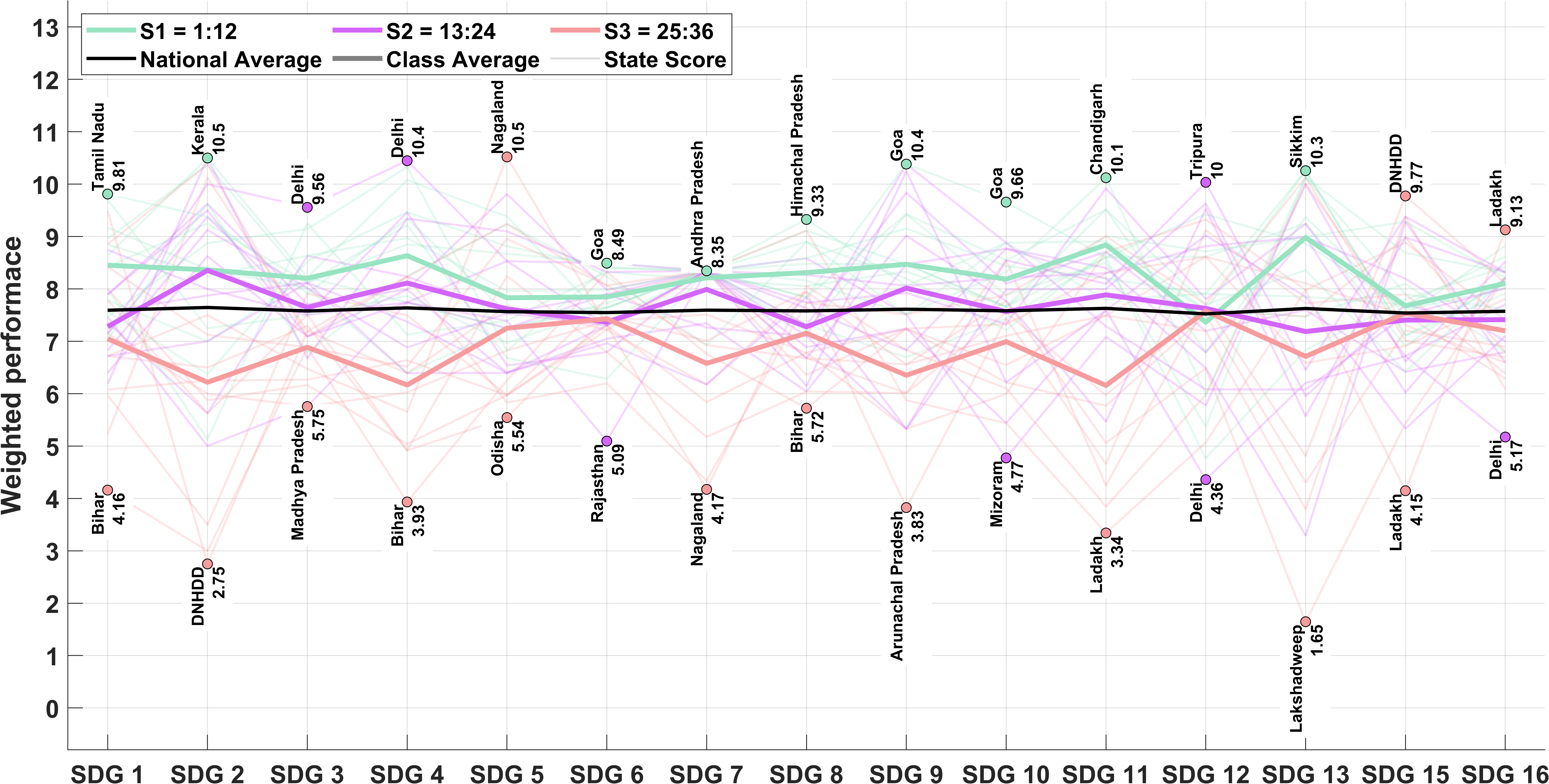}
\caption{\textbf{Weighted performance of states across goals.} After multiplying the weights of the goals that we have calculated with the performance of each state in each goal as $I_{sg}(2024)*W_g$ and the new weighted performances of all states are plotted. For comparison, the states are divided into 3 groups based on their NITI-Aayog ranks, and the average of their performances are also plotted as thicker lines along with the national average.}
\label{fig:Fig3}
\end{figure}

The vector $D_s(\tau)$ is used to rank the states and $C_g(\tau)$ is used to rank the sustainable development goals. The rankings of states obtained using the SDGs - GENEPY framework and the rankings from just the $k_s$ values are shown in Figure \ref{fig:Fig2}A and Figure \ref{fig:Fig2}B respectively where each of the states is colored according to these ranks. The scores for the SDGs obtained from the vector $C_g(\tau)$ is normalized $W_g(\tau)=C_g(\tau)/k'g(\tau)$ and plotted in a polar histogram as shown in Figure \ref{fig:Fig3}. As opposed to NITI-Aayog's equal weights when calculating the average scores of each state, these weights give the importance of each goal when calculating the SDGs-GENEPY complexities of states and their ranks. Since $D_s$ and $W_gs$ are calculated from the eigenvectors of a similarity matrix between the states and goals, we can look at it from a spectral clustering perspective\cite{bottai2024,BALLAND2022104450,INOUA2023104793}.

\begin{figure}[h]
\centering
\includegraphics[width=.45\linewidth]{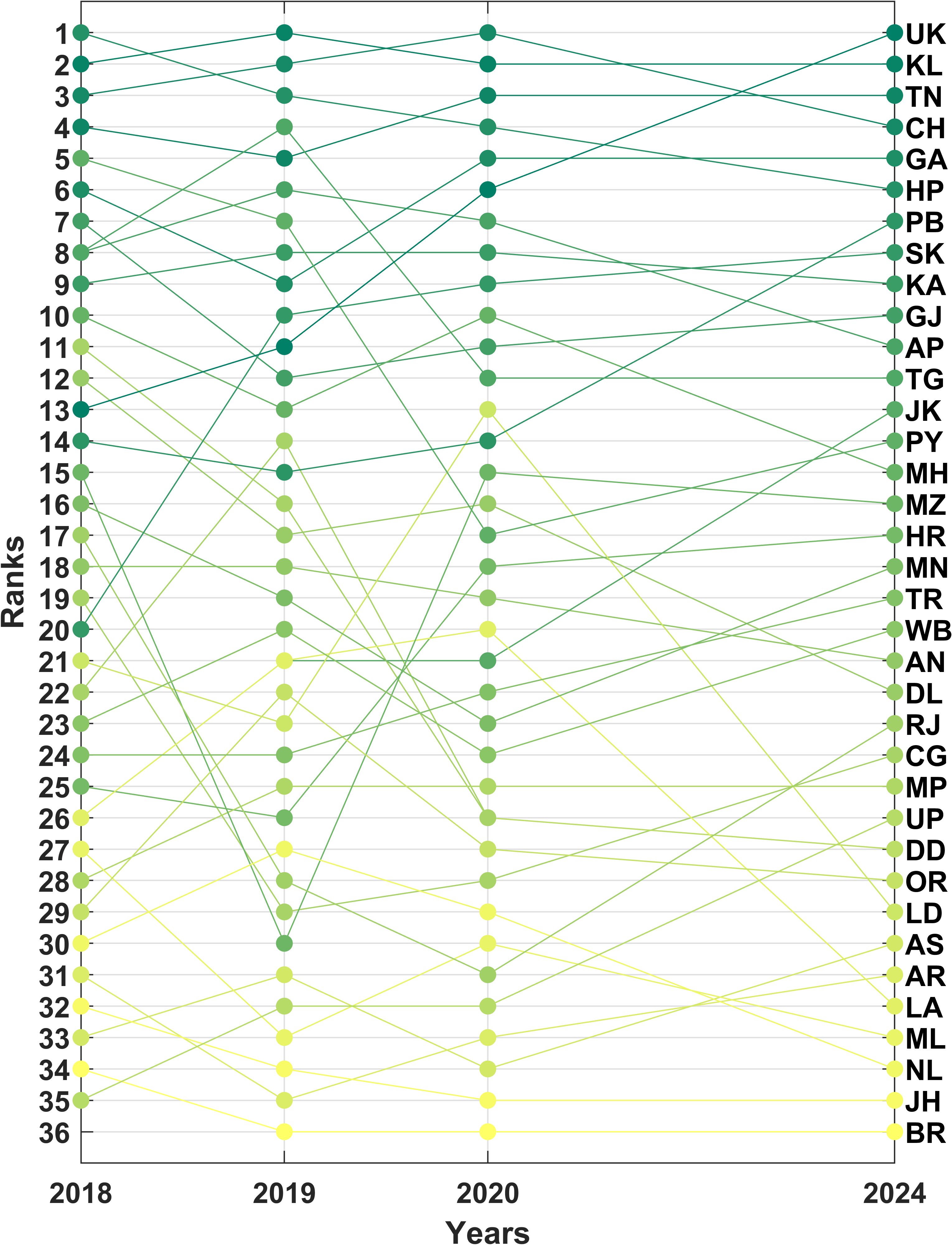}
\includegraphics[width=.45\linewidth]{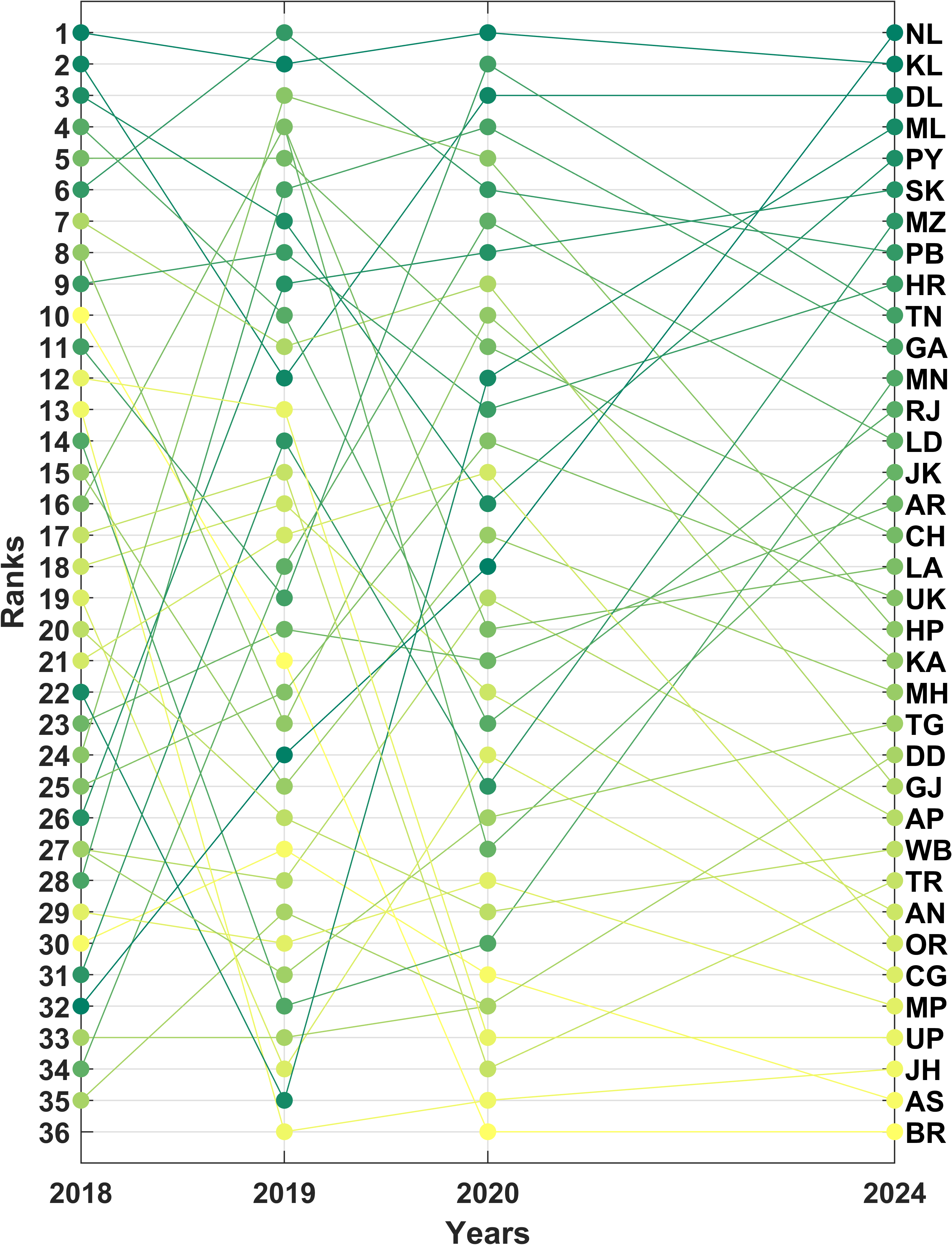}
\caption{\textbf{Evolution of ranks.} Evolutions of ranks of the states through the 4 years (2018,2019,2020,2024) of data obtained from $k_s(\tau)$ as well as $D_s(\tau)$ are plotted. Sub-figure A represents the evolution of $k_s(\tau)$ ranks (which are equivalent to the NITI-Aayog ranks) and sub-figure B shows the evolution of $D_s(\tau)$ ranks (SDGs-GENEPY ranks)}
\label{fig:Fig4}
\end{figure}

\begin{figure}[H]
\centering
\includegraphics[width=.9\linewidth]{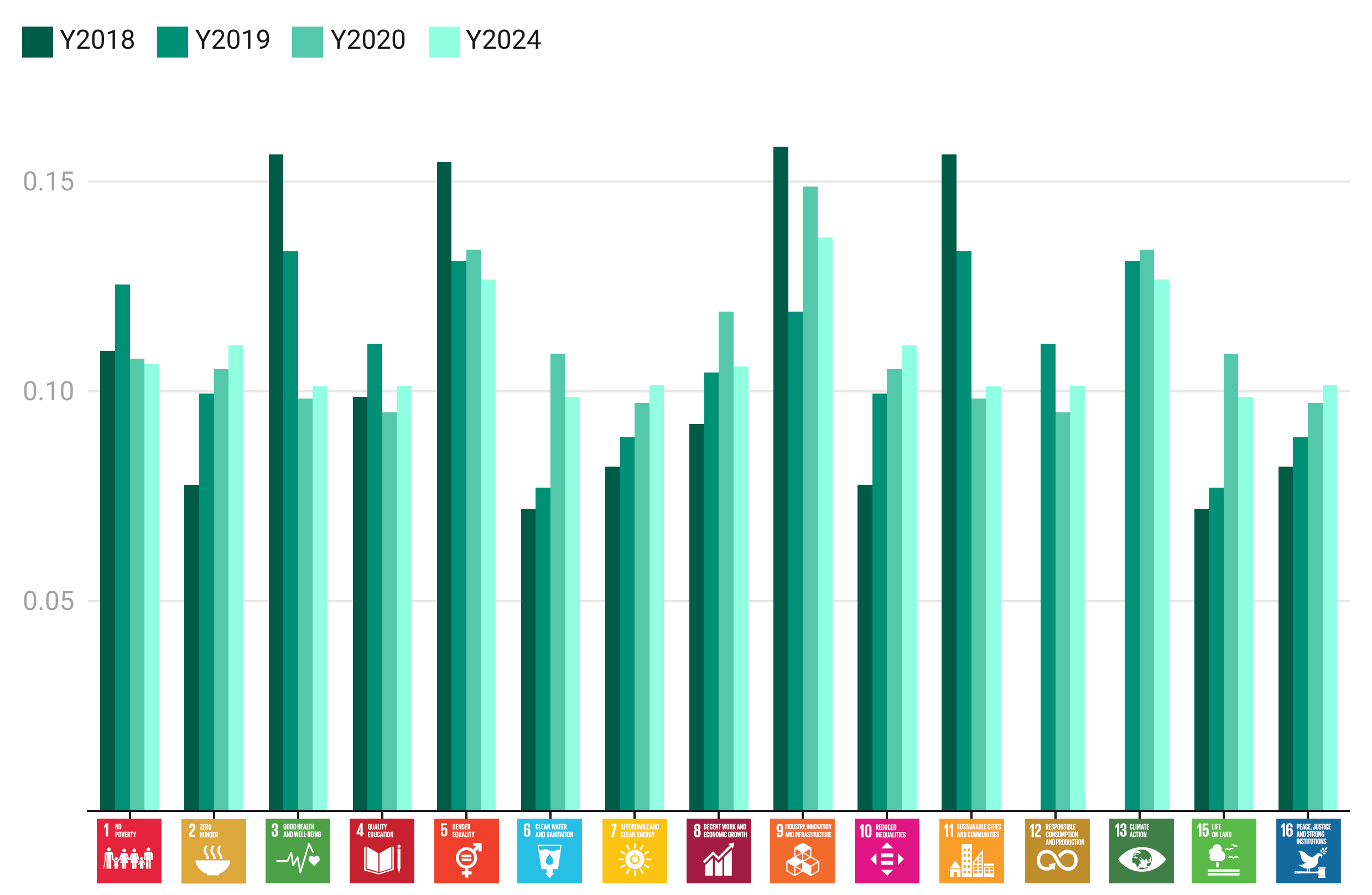}
\caption{\textbf{Evolution of weights.} The weights($W_g(\tau)$) of the goals obtained using SDGs-GENEPY framework over the four years (2018,2019,2020,2024) of data is shown as a grouped bar chart. Different groups of bars represent different SDGs and the colours of the bars in each group represent the year}
\label{fig:Fig5}
\end{figure}

\section*{Results and Discussion}
The ranks for the state/UTs as well as the weights of the SDGs are calculated from the centralities of the weighted bipartite network. Using a geographical plot, the ranks of the states are visualized in Figure \ref{fig:Fig2}a. The greener colors represent a higher rank for a state. A visual representation of the bi-partite network with a few selected states are shown in Figure \ref{fig:Fig2}b. These states are marked in the geographical plot for reference. The SDGs are indicated by their icons. The links (edges) between SDGs and states show the score NITI-Aayog gave that state on that goal. The colour as well as the thickness of the link indicates the score. The greener and thicker links show higher scores whereas the yellower and thinner links indicate lower scores. The weights calculated for each goal using SDGs-GENEPY formalism are shown as the horizontal bars near the corresponding SDG icons in Figure \ref{fig:Fig3}c. The values of these weights are also given as text in each bar.

Using the weights obtained from the formalism, we can rescale the scores of each state across goals to get a weighted performance ($I_{sg}(2024)*W_g$). These weighted performances of all the states and UTs for 2024 are plotted in Figure \ref{fig:Fig3} as line graphs. The x-axis corresponds to each SDG, and the y-axis has the weighted performance of a state in the corresponding SDG. Based on their $K_c$ ranks (similar to NITI-Aayog) the states are classified into 3 sets, each consisting of 12 states, and the line of each state is colored based on the set they belong to. The best and worst performing states in each of the SDGs are also highlighted along with their performance values. The average weighted performances of these 3 sets are also plotted in bold lines along with the national average in the corresponding colors.

Figure \ref{fig:Fig4} shows the evolution of the ranks of each state across the four years (2018,2019,2021 and 2024) obtained from both $K_s$ and $D_s$. Each state is represented by a colour based on its 2024 rankings. Even though the NITI-Aayog scores are equivalent to $K_s$ they had calculated separate rankings for states and UTs separately. For our study, we will be ranking them together which can create mismatches between the NITI-Aayog rankings and $K_s$ rankings. The number of states/UTs across years has changed because some states/UTs were newly made by splitting old states (Telengana and Ladak from Andhra Pradesh, Jammu and Kashmir), or merging some older ones (Dadra and Nagar Haveli and Daman and Diu from merging Dadra and Nagar Haveli and Daman and Diu). %What a confusing name for a UT

The evolution of weights($W_g(\tau)$) of the goals obtained using SDGs-GENEPY framework over the four years (2018, 2019, 2020, 2024) of data are shown as a grouped bar chart. Different groups of bars represent different SDGs and the colours of the bars in each group represent the year. For goals 12 and 13, the data was not available for the year 2018.

Concrete plans have been outlined globally to achieve the Sustainable Development Goals (SDGs) set under Agenda 2030, and India is following suit. Although there are various composite indices to measure SDG attainment worldwide, there is a lack of state-level analysis in India, necessitating a complex approach to evaluate each state’s progress. This study addresses this gap by establishing a bipartite network that assesses the SDG status across Indian states and union territories and ranks them based on their contributions and current status concerning the SDGs. Using NITI Aayog’s rankings as a benchmark, the study provides a basis for comparison and highlights that while there is a positive correlation between this study’s index and the NITI Aayog rankings, the relationship remains weak.

The study finds that SDG 9—focused on Industry, Innovation, and Infrastructure—ranks the highest across Indian states, signaling India’s emphasis on enhancing industry partnerships, improving business infrastructure, and fostering a culture of innovation nationwide. Achieving this goal significantly benefits the national economy, opens pathways for global industry exposure, and lays the foundation for international partnerships. Other important goals identified include SDG 2 (Zero Hunger), SDG 5 (Gender Equality), and SDG 13 (Climate Action). The analysis also highlights the top six states leading in SDG attainment, with Kerala ranking first due to its robust development strategies, such as the use of clean fuel in households, comprehensive sanitation measures, and high life expectancy coupled with the lowest infant mortality rate among states.

This research not only offers a holistic evaluation of states’ performance but also provides a relative assessment of their achievements and opportunities for improvement. The approach incorporates various dimensions and allows for a deeper understanding of challenges and opportunities specific to each state, offering significant managerial and policy implications for sustainable development strategies.

In conclusion, the study acknowledges the complexity of achieving the Agenda 2030 goals and emphasizes the need for a comprehensive analysis using advanced complexity mapping models. For developing nations like India, accelerating efforts towards sustainability is critical. The study introduces an SDGs-GENEPY framework to serve as a reference for future research, potentially using more temporally consistent datasets or investigating detailed SDG sub-goals. This framework not only provides a bird’s-eye view of SDG attainment but also encourages further exploration into specific areas, aiming for a deeper and more refined understanding of sustainable development progress across Indian states.

\section*{Summary and Remarks}
The results of this study offer a nuanced view of state and UT rankings across India in relation to their Sustainable Development Goal (SDG) achievements, utilizing a weighted bipartite network approach. The visual representation of state ranks in Figure \ref{fig:Fig2}a, with greener colors denoting higher rankings, captures regional disparities, allowing us to identify which states excel and which lag in SDG performance. By mapping selected states' performances across specific SDGs, the bipartite network in Figure \ref{fig:Fig2}b illustrates each state's strength on individual goals, where thicker, greener links reflect high scores, indicating progress and alignment with development goals. Conversely, thinner, yellower links indicate areas needing targeted interventions.

The weights of SDGs, as computed using the SDGs-GENEPY framework and visualized in Figure \ref{fig:Fig3}c, show the relative significance of each goal in the composite index. This weighted approach refines state scores by emphasizing SDGs with greater developmental impact, which is visually captured through the line graphs in Figure \ref{fig:Fig3}. Here, weighted performances highlight high-performing states, uncovering areas where development momentum is strongest. Classifying states into three performance sets illustrates the disparities between states while providing a national benchmark for SDG progress, allowing policymakers to target low-performing states effectively.

The progression of ranks across four years, displayed in Figure \ref{fig:Fig4}, provides insights into developmental shifts and how new administrative divisions, like Telangana and Ladakh, impact SDG outcomes. Ranking adjustments underscore regional developments and state responses to policy changes over time. Additionally, the annual evolution of SDG weights, depicted in a grouped bar chart, indicates shifting priorities, with notable data gaps for goals 12 and 13 in earlier years, underscoring the need for consistent, comprehensive data collection.

These findings emphasize that state-specific strategies informed by weighted SDG assessments can significantly enhance resource allocation and policy precision, enabling a more targeted approach to achieve sustainable, balanced development across diverse regions.

\section*{Acknowledgements}

SDG acknowledges the research seed grant from Hero MotoCorp Ltd. given to BMU.

%\bibliography{athesis}

% \section*{Author contributions statement}

% Must include all authors, identified by initials, for example:
% A.A. conceived the experiment(s),  A.A. and B.A. conducted the experiment(s), C.A. and D.A. analysed the results.  All authors reviewed the manuscript. 

% \section*{Additional information}

% To include, in this order: \textbf{Accession codes} (where applicable); \textbf{Competing interests} (mandatory statement). 

% The corresponding author is responsible for submitting a \href{http://www.nature.com/srep/policies/index.html#competing}{competing interests statement} on behalf of all authors of the paper. This statement must be included in the submitted article file.

% Figures and tables can be referenced in LaTeX using the ref command, e.g. Figure \ref{fig:stream} and Table \ref{tab:example}.

\end{document}